\documentclass[fleqn,usenatbib]{mnras}

\usepackage{newtxtext,newtxmath}
\usepackage[T1]{fontenc}

\usepackage{graphicx}	% Including figure files
\usepackage{amsmath}	% Advanced maths commands
\title[A Subgrid Turbulent Dynamo Model]{A Subgrid Turbulent Mean Field Dynamo Model for Cosmological Galaxy Formation Simulations}

\author[Y. Liu et al.]{
Yuankang Liu,$^{1,2,4}$\thanks{E-mail: yuankang.liu@durham.ac.uk}
Michael Kretschmer$^{1}$
and
Romain Teyssier$^{1,3}$\thanks{E-mail: teyssier@princeton.edu}
\\
$^{1}$Institute for Computational Science, Universität Zürich, Winterthurerstrasse 190, CH-8057 Zürich, Switzerland\\
$^{2}$Department of Physics, Eidgenössische Technische Hochschule Zürich, Otto-Stern-Weg 1, CH-8093 Zürich, Switzerland\\
$^{3}$Department of Astrophysical Sciences, Princeton University, 4 Ivy Lane, Princeton, New Jersey, 08544, United States\\
$^{4}$Institute for Computational Cosmology, Department of Physics, University of Durham, South Road, Durham DH1 3LE, UK
}

\date{Accepted XXX. Received YYY; in original form ZZZ}

\pubyear{2020}

\begin{document}
\label{firstpage}
\pagerange{\pageref{firstpage}--\pageref{lastpage}}
\maketitle

% Abstract of the paper
\begin{abstract}
Magnetic fields have been included in cosmological simulations of galaxy formation only recently. In this paper, we develop a new subgrid model for the turbulent dynamo that takes place in the supersonic interstellar medium in star-forming galaxies. It is based on a mean-field approach that computes the turbulent kinetic energy at unresolved scales and modifies the induction equation to account for the corresponding $\alpha$ dynamo. Our subgrid model depends on one free parameter, the quenching parameter, that controls the saturation of the subgrid dynamo. Thanks to this mean-field approach, we can now model the fast amplification of the magnetic field inside turbulent star-forming galaxies without using prohibitively expensive high-resolution simulations. We show that the evolution of the magnetic field in our zoom-in Milky Way-like galaxy is consistent with a simple picture, in  which the field is in equipartition with the turbulent kinetic energy inside the star-forming disc, with a field strength around 10~$\mu$G at low redshift, while at the same time strong galactic outflows fill the halo with a slightly weaker magnetic field, whose strength (10~nG) is consistent will the ideal MHD dilution factor. Our results are in good agreement with recent theoretical and numerical predictions. We also compare our simulation with Faraday depth observations at both low and high redshift, seeing overall good agreement with some caveats. Our model naturally predicts stronger magnetic fields at high redshift (around 100~$\mu$G in the galaxy and 1~$\mu$G in the halo), but also stronger depolarisation effects due to stronger turbulence at early time.
\end{abstract}

% Select between one and six entries from the list of approved keywords.
% Don't make up new ones.
\begin{keywords}
methods -- numerical: galaxy -- formation: galaxy -- magnetic fields.
\end{keywords}

%%%%%%%%%%%%%%%%%%%%%%%%%%%%%%%%%%%%%%%%%%%%%%%%%%

%%%%%%%%%%%%%%%%% BODY OF PAPER %%%%%%%%%%%%%%%%%%

\section{Introduction}

The origin of magnetic fields in the Universe remains an open question in modern cosmology. 
Using Faraday rotation measurements and synchrotron emission, it is possible to estimate the 
field strength in nearby, present-day galaxies at roughly a few to a few hundreds of $\mu$G \citep{Beck.1996,Beck.2015}. Using quasars absorption spectra, it is also possible to estimate the field strength in more distant galaxies \citep{Widrow.2002}. Surprisingly, the field strength appears higher in the past than it is today \citep{Bernet.2008, Krause.2009o3q, Kim.2016dx2}. This poses severe constraints on theoretical models, as any underlying magnetic dynamo mechanism must be very fast to accommodate for these observations.

Traditional galactic dynamo theories are based on the large scale dynamo model \citep{Larmor.1919, Cowling.1933, Parker.1955, Parker.1970, Brandenburg.2005}, where small-scale helical motions work in tandem with large-scale rotation to drive the amplification of the field. These models are justified for present-day galaxies dominated by thin and quiescent spiral discs like our own Milky Way. These large-scale dynamo models explain many of the detailed observational properties of nearby galactic magnetic fields. They are however not fast enough to account for the strong magnetic fields observed at high-redshift in the direction of distant quasars.

In addition, our understanding of the galaxy formation process has evolved quite dramatically in the past decade, strengthening the role of stellar feedback processes in driving strong galactic outflows \citep{Oppenheimer.2006, Scannapieco.2012}. This change in the paradigm was motivated by the ubiquitous observation of strong outflows in high-redshift galaxies \citep{Steidel.2010, Schreiber.2014} and their very disturbed morphologies. As a consequence, we believe now that high-redshift gas discs are very turbulent and thick, for which the large scale dynamo picture does not apply. 

Many recent theoretical and numerical studies have shown that high-redshift galaxies are in fact subject to a strong turbulent dynamo \citep{Beck.2012, Rieder.2016, Pakmor.2017, Butsky.2017, Pakmor.2020, Steinwandel.2018, Steinwandel.2020hfw}. It is sometimes called the small-scale dynamo because it amplifies the magnetic field only at the injection scale of turbulence and below, while the large-scale dynamo can amplify the field on larger scales. Note that in most cases, the injection scale of turbulence is close to the disc thickness. As a consequence, this scale is quite small at low redshift, for which the typical disc thickness is around 100~pc, and it is quite large at high redshift, where even dwarf galaxies have a disc thickness as large as 1~kpc.

In \cite{Rieder.2016}, we have simulated this turbulent dynamo in an isolated high-redshift dwarf galaxy. We have shown that indeed strong supernova-driven outflows can drive a fast turbulent dynamo as soon as the spatial resolution is high enough. It is a well-known property of the turbulent dynamo that the dissipation scale has to be smaller than the kinetic energy injection scale by a factor of at least 30. In other words, in order to power a magnetic dynamo, the flow Reynolds number has to be larger than a critical value estimated numerically around 30-60 \citep{Brandenburg.2005}. 

In the context of numerical simulations, dissipation is provided by local truncation errors, translating this into a resolution requirement. Assuming a disc thickness of 1~kpc, this means we need a resolution of at least 30~pc to see a turbulent dynamo. This is roughly what was observed both in the idealized numerical experiments of \cite{Rieder.2016} and in a zoom-in cosmological simulation in \cite{Rieder.2017}. This is bad news for cosmological magneto-hydrodynamics. Although a spatial resolution of 20~pc for thick galactic discs is not impossible on modern architectures, the situation is more critical for low redshift, razor thin discs. Indeed, in this case, with a disc thickness of 100~pc, we obtain the much more severe resolution requirement of 3~pc.

Note that many past MHD simulations of galaxy and galaxy cluster formation suffer from this critical resolution problem \citep{Dolag.2002, Dubois.2010, Vazza.2014, Rieder.2017, Martin-Alvarez.2018}. The consequence of having a resolution that is too low, is to obtain unrealistically a very slow magnetic dynamo, if any \citep[but see][for recent simulations at much higher resolution with faster dynamos]{Vazza.2017, Steinwandel.2021}. In order to reach equipartition between the magnetic energy and the other important energy densities, one has to start with a high enough seed field.  Unfortunately, if this initial seed field is too strong, it has also unrealistically a strong impact on the collapse of the first objects and spuriously affect their dynamics. The resulting fine tuning of the adopted initial conditions is quite unsatisfactory. 

The most popular model to explain the origin of seed magnetic fields in the Universe is the Biermann battery process \citep{Biermann.1950}. Small velocity drifts between electrons and ions in presence of misaligned electron density and electron pressure gradients generate microscopic currents and fields during the Epoch of Reionization at the level of $10^{-20}$~G \citep{Kulsrud.1997, Gnedin.2000, Garaldi.2021, Attia.2021}. This is very far from the observed $10^{-6}$~G observed in galaxies today and even $10^{-4}$~G in the past. We therefore need a very vigorous 
turbulent dynamo and cannot afford these limitations.

This motivates the design of a new subgrid turbulent dynamo model that can overcome these problems.
In the context of turbulence, the mean-field  theory provides an efficient framework to divide the turbulence scales into: 1- the resolved scales where turbulent processes can be captured by the hydrodynamics solvers, defining thus the mean field by the numerical solution, and 2- the unresolved scales, defining the fluctuations with respect to the mean that can only be accounted for by a subgrid model. Such a model has been used to follow unresolved turbulence in galaxy formation simulations \citep{Braun.2014, Semenov.2016, Kretschmer.2019} and to set the local star formation efficiency. It can also be exploited to predict the line emission of the CO molecule from the unresolved, dense molecular gas \citep{Kretschmer.2021}. 

In this paper, we will exploit our subgrid scale (SGS) model for the unresolved turbulence to inform a mean-field model for the induction equation, allowing us to describe a turbulent dynamo, even if we lack the necessary spatial resolution. The paper is organised as follows: In section~\ref{sec:num}, we describe the numerical methods, in particular our subgrid turbulent dynamo model. In section~\ref{sec:Faraday}, we describe the methodology we used to compare our results to observations of Faraday rotation. In section~\ref{sec:cosmo}, we describe in more detail the cosmological zoom-in simulation of a Milky Way analogue that will be used in section~\ref{sec:results} to analyze the amplification process and the properties of the magnetic field both at high and low redshift. 
Finally, we discuss the implications of our results in section~\ref{sec:discussion}. 

\section{Numerical Methods}
\label{sec:num}

We model the amplification of an initially weak and constant seed field using a cosmological zoom-in simulation of a Milky Way-like galaxy performed by the Adaptive Mesh Refinement (AMR) code RAMSES \citep[][]{2002A&A...385..337T}. The simulation contains a collisionless fluid (made of dark matter and stars) and a magnetised gaseous component, coupled through gravity. Beyond the traditional cooling and UV heating of the gas, we model star formation and stellar feedback using standard subgrid models. In this section, we first describe the numerical methods for solving the ideal MHD equations. Then we introduce our subgrid turbulent dynamo model for simulating the amplification and evolution of magnetic fields across cosmic time. Finally, we list our various parameters used in our galaxy formation model. 

\subsection{Ideal MHD Solver}
\label{sec:numerical_methods} % used for referring to this section from elsewhere

The dynamics of the baryonic component is described using the equations of ideal MHD, written in conservative form as: 
\begin{align}
    \label{eq:MHD_1}
    \partial_t \rho + \nabla \cdot (\rho \mathbf{u}) & = 0 \\ 
    \label{eq:MHD_2}
    \partial_t (\rho \mathbf{u}) + \nabla \cdot (\rho \mathbf{u} \mathbf{u}^\mathrm{T} - \mathbf{B} \mathbf{B}^\mathrm{T} + P_{\rm tot}) & = \rho \mathbf{g} \\
    \label{eq:MHD_3}
    \partial_t E + \nabla \cdot [(E + P_{tot}) \mathbf{u} - (\mathbf{u} \cdot \mathbf{B})\mathbf{B}] & =  \Gamma - \Lambda + \rho \mathbf{g} \cdot \mathbf{u}\\
    \label{eq:MHD_4}
    \partial_t \mathbf{B} - \nabla \times (\mathbf{u} \times \mathbf{B}) & = 0,
\end{align}
where $\rho$ is the gas density, $\mathbf{g}$ the gravitational acceleration, $\rho \mathbf{u}$ the momentum, $\mathbf{B}$ the magnetic field, $E = \frac{1}{2} \rho \mathbf{u}^2 + \rho \varepsilon + \frac{1}{2}\mathbf{B}^2$ the total energy, $\Gamma$ the heating function, $\Lambda$ the cooling function and $\varepsilon$ is the specific internal energy. $P_{\rm tot}$ is the total pressure given by $P_{\rm tot} = P + \frac{1}{2}\mathbf{B}^2$. We solve the ideal MHD system along with a perfect gas equation of state:
\begin{equation}
\label{EoS}
P = (\gamma - 1) \rho \varepsilon,
\end{equation}
where $\gamma$ is the adiabatic index. We also need to augment the system with the solenoidal constraint 
\begin{equation}
\label{solenoidal}
\nabla \cdot \mathbf{B} = 0.
\end{equation}
The four ideal MHD equations are solved using the second-order unsplit Godunov scheme \citep[][]{2006A&A...457..371F} based on the MUSCL-Hancock scheme. The induction equation (eq.~\ref{eq:MHD_4}) is solved using the Constrained Transport (CT) method \citep[][]{2006JCoPh.218...44T}, which keeps the divergence of the magnetic field $\nabla \cdot \mathbf{B} = 0$ down to machine precision.

\subsection{Subgrid Turbulent Dynamo Model}
\label{sec:subgrid_turbulent_dynamo_model}

We now describe our model for the subgrid turbulent dynamo. It is based on the so-called mean-field approach of subgrid turbulence \citep[see e.g.][and reference therein]{Schmidt.2011}, but applied to the MHD equations. Mean-field electrodynamics was first formalized over half a century ago~\citep{1966ZNatA..21..369S, Parker.1970, Ruzmaikin.1988} and has been used in dynamo modelling ever since. Mean-field electrodynamics relies on a scale separation between fluctuating and mean quantities. We thus consider the decomposition 
\begin{equation}
\label{eq:decomposition}
\mathbf{u} = \overline{\mathbf{u}} + \mathbf{u}',\quad \mathbf{B} = \overline{\mathbf{B}} + \mathbf{b}',
\end{equation}
where $\overline{\mathbf{u}}$ and $\overline{\mathbf{B}}$ are the mean velocity and magnetic fields (resolved by the grid), while $\mathbf{u}'$ and $\mathbf{b}'$ are the fluctuations (unresolved by the grid). Substituting Eq.~\ref{eq:decomposition} into Eq.~\ref{eq:MHD_4} and averaging, we obtain the mean-field induction equation,
\begin{equation}
\label{eq:mean-field_induction_eq}
\frac{\partial \, \overline{\mathbf{B}}}{\partial t} = \nabla \times (\overline{\mathbf{u}} \times \overline{\mathbf{B}} + \overline{\mathcal{E}}),
\end{equation}
where
\begin{equation}
\label{eq:emf_eq}
\overline{\mathcal{E}} = \overline{\mathbf{u}' \times \mathbf{b}'}
\end{equation}
is the electromotive force (EMF) corresponding to the fluctuations only. 
Identifying the original MHD equation with the mean-field terms only, we see that this new EMF act as a new source term for the evolution of $\overline{\mathbf{B}}$. Unfortunately, we need an expression for $\overline{\mathcal{E}}$ in terms of the mean field $\overline{\mathbf{B}}$, which is a standard closure problem at the heart of mean-field theory~\citep{Brandenburg.2005}. 

In the two-scale approach~\citep{1978mfge.book.....M}, one assumes that $\overline{\mathcal{E}}$ can be expanded in powers of the gradients of the mean magnetic field, which can be written as 
\begin{equation}
\label{eq:emf_expression}
\mathcal{E}_i = \alpha_{ij} \overline{B}_j + \beta_{ijk} \frac{\partial}{\partial \, x_k}\overline{B}_j + ...,
\end{equation}
where $\alpha_{ij}$ and $\beta_{ijk}$ are known as the turbulent transport coefficients. They depend on the stratification, angular velocity, etc. The simplest case, which still captures the essence of the problem, is that Eq.~\ref{eq:emf_expression} is truncated after the first term
\begin{equation}
\label{eq:emf_expression_1st_order}
\mathcal{E}_i = \alpha_{ij} \overline{B}_j.
\end{equation}
If the turbulence is homogeneous and isotropic, the tensor $\alpha_{ij}$ takes the form of $\alpha_{ij} = \alpha \delta_{ij}$ where $\alpha$ is a scalar and Eq.~\ref{eq:emf_eq} becomes
\begin{equation}
\label{eq:emf_expression_simplified}
\overline{\mathcal{E}} = \alpha \overline{\mathbf{B}}.
\end{equation}
Under certain conditions, $\alpha$ can be estimated using the average subgrid kinetic helicity and the correlation time scale of turbulent eddies as
\begin{equation}
\alpha = -\frac{{\tau}_{\rm corr}}{3} \overline{ {\bf u} \cdot ( \nabla \times {\bf u}) }.
\end{equation}
This contribution of this term to the mean-field induction equation is known as the $\alpha$-effect in many dynamo models. Determining the exact value for $\alpha$ is a field of research of its own  \citep{1996PhRvE..54.4532C, Brandenburg.2005, 2007GApFD.101..117R, 2009MNRAS.395L..48C, 2008A&A...486L..35G, 10.1093/mnras/sts356}.
Some models propose a new mean-field equation to describe the time evolution of $\alpha$, other models adopt fixed analytical closed forms. All these models are highly controversial and heavily debated in the literature.

Here, we propose the following simple model for $\alpha$, based on the known value of the SubGrid Scale (SGS) (see Sec.~\ref{sec:subgrid_models_galaxy}) turbulent velocity dispersion $\sigma_\mathrm{T}$ 
\begin{equation}
\label{eq:alpha_expression}
\alpha_\mathrm{T} = \sigma_\mathrm{T} 
\max \left[ 1 - \frac{E_{\mathrm{mag}}}{q K_{\rm T}},0 \right].
\end{equation}
Because it depends on the unresolved turbulent velocity dispersion in the subgrid turbulence model that will be described in Section~\ref{sec:subgrid_models_galaxy}, we use the index ${\rm T}$ to outline its turbulent origin. In the previous equation, $E_{\mathrm{mag}}$ is the mean-field magnetic energy density defined by $E_{\rm mag} = \frac{\overline{B}^2}{8\pi}$ and the turbulent kinetic energy density, defined by $K_{\rm T}=\frac{3}{2}\rho \sigma_\mathrm{T}^2$. We introduce a free parameter $q$ that controls the quenching of the subgrid turbulent dynamo. Indeed, one can see easily that once the magnetic energy density reaches a fraction of the turbulent kinetic energy, $\alpha_{\rm T}$ goes strictly to zero and the subgrid turbulent dynamo vanishes. 

The quenching parameter $q$ is the key parameter of our subgrid turbulent dynamo.
It is closely related to the saturation of the turbulent dynamo when the magnetic energy approaches the
turbulent kinetic energy and the back reaction of the Lorentz force on the turbulent flow effectively stops the dynamo. Based on an effective turbulent resistivity determined by the turnover time scale of turbulent eddies and the magnetic energy density, one can come up with a scale-dependent saturation model and estimate the field strength at saturation analytically~\citep{PhysRevE.92.023010}. The saturated field strength depends on the magnetic Reynolds number and the type of turbulence, and it differs in the limits of large and small magnetic Prandtl numbers~\citep{PhysRevE.92.023010}. The ratio between the saturated magnetic energy density and the turbulent kinetic energy usually ranges from 0.1\% to 50\%~\citep{PhysRevE.92.023010}. 

Note that in our model, the subgrid turbulent dynamo only augments towards smaller, unresolved scales the marginally resolved turbulent dynamo in the simulation. It is therefore important to choose a relatively low value of the quenching parameter so that the resolved flow is able to take over and bring the field to higher strength, possibly to equipartition. We explored various values of the quenching parameter, as seen in Fig.~\ref{fig:epsilon_effect}, and decided to choose $q=10^{-3}$ as our fiducial model (see more discussion later). 

In order to restrict the subgrid turbulent dynamo to the supersonic ISM in the galactic discs, we only allow $\alpha_{\rm T}$ to be non-zero in star forming gas. We therefore use a density threshold for the subgrid dynamo identical to the density threshold for turbulent star forming gas, namely $n_{\rm H} > 10^{-2}$~H/cc, as explained in Sec.~\ref{sec:subgrid_models_galaxy}. This has an important consequence for the growth of the magnetic field in the intergalactic medium. In our model, there is no subgrid turbulent dynamo in the diffuse hot halo gas. This means that in our simulations, strong magnetic fields {\it outside of galactic discs} can only come from the pollution of galactic outflows, and not from in situ (generally subsonic) turbulence. Note however that with a higher resolution in the CGM , we would probably see an additional dynamo amplification of the magnetic field in the low density medium, as reported for example in \cite{Vazza.2017} and \cite{Steinwandel.2021}.

\subsection{Subgrid Models for Galaxy Formation Physics}
\label{sec:subgrid_models_galaxy}

We model the unresolved turbulence using the SGS model  proposed recently to describe turbulent effects at the macroscopic scale \citep{2005CTM.....9..693S, Schmidt.2006, 2011A&A...528A.106S}. We introduce a new equation for the turbulent kinetic energy $K_\mathrm{T}$ \citep[][]{2014nmat.book.....S, Semenov.2016, 2020MNRAS.492.1385K}, evolving it alongside the other MHD conservative variables. In this formalism, turbulence arises from the competition between a creation term due to large scale shear and a destruction term modelling the dissipation within the turbulent cascade down to viscous scales \citep[see][for more details]{Schmidt.2006,Semenov.2016,2020MNRAS.492.1385K}. 
The turbulent velocity dispersion $\sigma_{\rm T}$ can be calculated from the turbulent kinetic energy by $K_\mathrm{T}= 3/2 \rho \sigma_\mathrm{T}^2$.

Another important subgrid model for galaxy formation simulation is the star formation recipe. 
As many other past studies in the literature, we use the so-called Schmidt law, for which  the star formation rate density is defined as $\dot{\rho}_\star=\epsilon_{\rm ff} {\rho}/{t_\mathrm{ff}}$,
where $\rho$ is the gas density, $t_\mathrm{ff}=\sqrt{{3\pi}/{(32 G \rho)}}$ is the local gas free-fall time and $\epsilon_{\rm ff}$ is the star formation efficiency per free-fall time.
Traditionally, $\epsilon_{\rm ff}$ is chosen to be a constant usually close to $1\%$. In addition, it is required that the gas density is above a certain (resolution dependent) density threshold. Additional criteria can be added such as the gas temperature being cold enough or the flow being converging.

In our approach, $\epsilon_{\rm ff}$ is based on the turbulent state of the gas within the cell. This is precisely the information that our SGS turbulence model can give us. Integrating over the unresolved lognormal density PDF, one can ask what is the fraction of unresolved molecular cores that are gravitationally unstable and derive the star formation rate for the entire computational cell.
It depends on the cell virial parameter $\alpha_\mathrm{vir}$ and the turbulent Mach number $\mathcal{M}$, both depending on $\sigma_{\rm T}$ \citep{Schmidt.2006, Semenov.2016, Kretschmer.2019}. Because unresolved turbulence can vary from cell to cell, this method allows for varying efficiencies that can span a large range of values $\epsilon_{\rm ff} = 0 - 100\%$. This model is capable of producing both very high star-formation efficiencies, typically during starbursts, as well as very low efficiencies in quenched galaxies \citep[see discussion in][]{Kretschmer.2019}. This subgrid turbulent star formation recipe is restricted to cells denser than $n_{\rm H}>10^{-2}$~H/cc. This density threshold corresponds to the self-shielded gas (see below) that can cool and develop sustained supersonic turbulence \citep[see][for a more in-depth discussion]{Schaye.2004}. 

The last important ingredient of our simulation is the implementation of stellar feedback. In our model, individual supernova explosions inject thermal energy of $E_\mathrm{SN}=10^{51} \mathrm{erg}$ into the surrounding gas {\it only if the local cooling radius of the Sedov blast wave is resolved}.
If the cooling radius is unresolved, which occurs usually at high gas densities, we additionally inject the terminal momentum of the blast wave in its snowplow phase into the surrounding cells \citep{Martizzi.2015}. Note that our star particles are massive enough to spawn multiple (typically  $\sim 1000$) supernova explosions between $3$ Myrs and $20$ Myrs after its birth.

In addition to these subgrid models, we use the traditional physics for galaxy formation simulations, which include: metal ejection from supernovae, equilibrium H  and He cooling and UV heating, metal cooling, self-shielding of the gas at densities larger than $n_{\rm H}>10^{-2}$~H/cc \citep{Aubert.2010, Kretschmer.2019, Agertz.2019}.

\section{Faraday Rotation Synthesis}
\label{sec:Faraday}

Faraday rotation corresponds to the rotation of the plane of polarisation of electromagnetic radiation propagating through a magnetised plasma \citep{burke_graham-smith_wilkinson_2019}. The Rotation Measure (RM) is defined as the ratio between the gradient of the polarisation angle $\chi$ and the wavelength $\lambda^2$:
\begin{equation}
    \mathrm{RM} = \frac{\Delta \chi}{\Delta \lambda^2}.
    \label{eq:rm}
\end{equation}
If all the radiation received in the light beam undergoes the same Faraday rotation, the gradient RM is equal to the Faraday depth $\phi$, which is usually given by \citep[][]{1966MNRAS.133...67B}
\begin{equation}
    \phi(z_s) = 8.1 \times 10^5 \int^0_{z_s} \frac{n_e(z)B_{||}(z)}{(1+z)^2}\frac{\mathrm{d}l}{\mathrm{d}z}\mathrm{d}z.
    \label{eq:Faraday_Depth_eq}
\end{equation}
Here, $\phi$ is in units of rad m$^{-2}$, the free electron density $n_e$ in units of cm$^{-3}$, the magnetic field projection along line of sight $B_{||}$ in units of G, and the patch increment per redshift $\frac{\mathrm{d}l}{\mathrm{d}z}$ is in units of pc. $z_s$ corresponds to the redshift of the background source.

There are cases where mixing of emission at different Faraday rotations occurs. One scenario is there may be several sources along the line of sight. Moreover, real radio telescopes have finite spatial resolutions. Inhomogeneous Faraday depth screens within the telescope beam could cause that the emissions at different Faraday depths are smeared indistinguishably. In such cases, the overall polarisation is reduced, which is often referred to as ``Faraday depolarisation'' \citep[][]{1966MNRAS.133...67B, 1966ARA&A...4..245G, 1998MNRAS.299..189S}. The depolarisation effect can lead to nonlinearities in the slope of $\Delta \chi$ against $\Delta \lambda^2$ and make the gradient RM poorly defined. 

The collected emission by real telescopes has been subject to a distribution of Faraday depth, denoted by $F(\phi)$, where the only exception is in the case of a uniform Faraday screen. Fortunately, $F(\phi)$ can be estimated from radio polarisation measurements using the Faraday Rotation Measure Synthesis (RM Synthesis) technique. Faraday RM synthesis transforms the complex polarisation representation
\begin{equation}
    \mathbf{P}(\lambda^2) = P(\lambda^2) e^{2i\chi(\lambda^2)},
    \label{eq:polarisation_representation}
\end{equation}
where $P$ is the polarised flux and $\chi$ is the polarisation angle, into a complex Faraday depth distribution
\begin{equation}
    \mathbf{F}(\phi) = \frac{1}{\pi} \int^{\infty}_{-\infty} \mathbf{P}(\lambda^2) e^{-2i\phi \lambda^2} \mathrm{d} \lambda^2.
    \label{eq:complex:faraday_distribution}
\end{equation}
Here the complex $\mathbf{F}(\phi)$ can be expressed as 
\begin{equation}
    \mathbf{F}(\phi) = F(\phi)e^{2i\psi(\phi)},
\end{equation}
where $F(\phi)$ is the Faraday depth spectrum and its phase $\psi(\phi)$ is the initial phase of the polarisation. In the simple case of only one homogeneous intervener across the plane of sight, $F(\phi)$ would have a single peak. The full $F(\phi)$ contains information about the structure of the intervening system.  

\citet{Kim.2016dx2} have used Faraday RM synthesis to obtain Faraday depth distributions and found correlation with Mg II absorption. They have found strong random fields of the order of 10 $\mu$G in the CGM. In their analysis, they used a number of parameters to characterise the structure of Faraday depth spectrum $F(\phi)$. Two of them are essential to our analysis: $\phi_{\rm max}$ and $\sigma_{\rm PC}$. $\phi_{\rm max}$ is the peak position of the Faraday depth distribution after removing the Galactic foreground contribution according to \citet{2015A&A...575A.118O} and $\sigma_{\rm PC}$ is obtained by fitting a Gaussian function to the primary component of the Faraday Depth distribution with the correction for the intrinsic spread function arising from the finite wavelength band-width of the radio data \citep[][]{Kim.2016dx2}. We will use these two important observables in the next sections to compare our simulations to existing Faraday rotation data.

\section{COSMOLOGICAL SIMULATION}
\label{sec:cosmo}

Our study is based on a cosmological zoom-in simulation that was performed with the adaptive mesh refinement code \texttt{RAMSES} \citep{2002A&A...385..337T}.
We first have run a low-resolution dark-matter-only N-body simulation with a box-size of $25$ Mpc$/h$ containing $512^3$ particles, using the standard $\Lambda$CDM cosmology parameters obtained by \cite{2020A&A...641A...6P}.
From this box we have selected several candidate halos, targeting objects that have a virial mass similar to the Milky Way, namely in the range $M_\mathrm{vir} = (0.5 - 1.5) \times 10^{12} \, \mathrm{M}_\odot$, where the virial mass was calculated according to the definition of \cite{Bryan.1998}.
Additionally, we require that these halos are in relative isolation at $z=0$ and that they accumulate their mass mostly before $z=1$ which excludes halos featuring late major mergers.

\subsection{Initial Conditions}

Using the MUSIC code \citep{Hahn.2011}, we have generated higher resolution initial conditions around the selected halos, where the initial grid levels ranged from $\ell_{\rm min}=7 $ to $\ell_{\rm max,ini}=11$.
The chosen maximum level for the initial grid $\ell_{\rm max,ini}=11$ corresponds to an effective initial resolution of $2048^3$, which yields a dark matter particle mass of $m_{\rm dm}=2.0 \times 10^{5} \, $M$_\odot$ and an initial baryonic mass of $m_{\rm bar}=2.9 \times 10^{4} \, $M$_\odot$. These haloes have been studied in great detail in a previous study \citep{Kretschmer.2020}. In the present paper, we use the halo that gave rise to a large, Milky Way-like disc and re-simulated it with MHD and all the above described subgrid models.

The maximum resolution during the course of the simulation was set to $\ell_{\rm max}=19$, with refinement levels progressively released to enforce a quasi-constant physical resolution.
We have used the traditional quasi-Lagrangian approach as refinement strategy, where cells are individually refined if more than 8 dark matter particles are present or if the total baryonic mass (gas and stars) exceeds $8\times m_{\rm bar}$. The highest resolution cells have sizes of $\Delta x_{\rm min}=55$pc, roughly constantly in physical units.
Only the Lagrangian volume corresponding to twice the final virial radius of the halo was adaptively refined, the rest of the box being kept at a fixed, coarser resolution to provide the proper tidal field.

The primordial magnetic seed field predicted by the linear perturbation theory of the Biermann battery is very weak, around $10^{-25}$~G \citep{2013PhRvL.111e1303N}. The non-linear evolution of the Biermann battery during cosmological ionization fronts gives rise to slightly higher field strength around $10^{-20}$ G \citep{2000ApJ...539..505G, Attia.2021}. To mimic these weak seed fields, we have initialised the magnetic field as a constant field parallel to the $z$-direction with a strength of $10^{-20}$ G in physical units. We postpone to a future paper the study of different and more complex initial magnetic field configurations. 

\subsection{Effects of the Quenching Parameter}

\begin{figure*}
    \centering
    \includegraphics[width=1.0\textwidth]{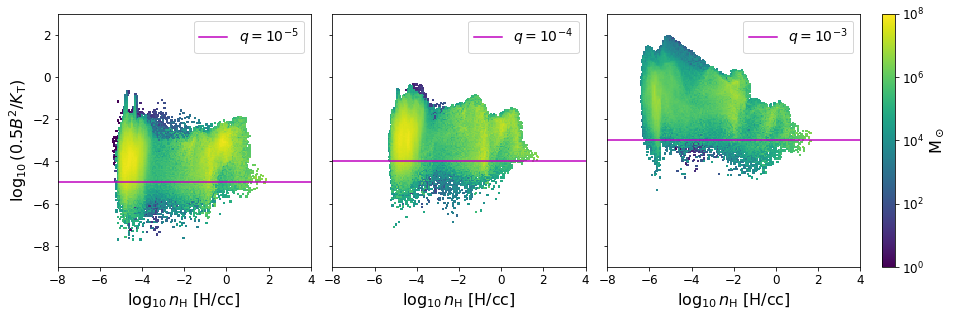}
    \caption{Mass weighted 2D histograms of the ratio of magnetic energy to turbulent kinetic energy versus gas density within $R_\mathrm{vir}$ at redshift $z = 0$. The three panels show three different simulations with different values of the quenching parameter $q = 10^{-5}$, $q = 10^{-4}$ and $q = 10^{-3}$ respectively. The straight horizontal lines indicate the ratio above which the subgrid turbulent dynamo is quenched (see Eq.~\ref{eq:alpha_expression}). The density threshold, $n_{\rm H}>10^{-2}$~H/cc as discussed in Sec.~\ref{sec:subgrid_turbulent_dynamo_model}, is kept the same in all three cases. }
    \label{fig:epsilon_effect}
\end{figure*}

In this section, we investigate the effect of varying the value of $q$, the quenching parameter of our subgrid $\alpha$ dynamo. For this, we have run a series of lower resolution MHD cosmological zoom-in simulations of the same galaxy. We show in Fig.~\ref{fig:epsilon_effect} 2D histograms of the ratio between the magnetic energy and the turbulent kinetic energy density versus the gas density for all the gas within $R_\mathrm{vir}$ at redshift $z=0$. At this late epoch, all the gas within the virial radius has been magnetised above the subgrid dynamo saturation value equal to $q$ (shown as the horizontal solid line in Fig.~\ref{fig:epsilon_effect}). The resolved motions (both the resolved turbulence and the global rotation in the disc) have managed in all cases to amplify the magnetic field 3 to 4 orders of magnitude further, but not much more, hence the need for a subgrid dynamo. 

With $q=10^{-3}$, we manage to reach equipartition for the simulated magnetic field, with a ratio of magnetic to turbulent kinetic energy equal or even larger than one. Although using larger values for $q$ is in principle possible, we argue against it as it would lead to unrealistically large field strengths close to equipartition everywhere. The free parameter $q$ is determined by experimenting different values. Note that the adopted value for $q$ is likely resolution dependent, as many $\alpha$ dynamo models suggest, with the quenching parameter $q$ being inversely proportional to the magnetic Reynolds number. The quench of dynamo will occur once the dynamo is a small fraction of unresolved turbulent kinetic energy, which decreases with numerical resolution. 

For the high resolution simulation, we have adopted the value $q=10^{-3}$ as our fiducial model.
We also see in Fig.~\ref{fig:epsilon_effect} that for this value, the lower density gas is significantly above equipartition with the turbulent energy. This is because in low density gas, the thermal energy is much larger than the turbulent kinetic energy. The magnetic field there is mostly originating from galactic outflows and not from the turbulent dynamo, in particular because we set $\alpha$ to be zero in gas less dense than $10^{-2}$~H/cc.

\section{Results}
\label{sec:results}

In this paper, we restrict our analysis to two different epochs: first at the rather high redshift $z=4$, with a typical example of a high-$z$ turbulent galaxy, and then at the relatively recent epoch $z=0.2$, with a typical example of a nearby quiescent and disc-dominated galaxy. In both cases, we have generated Faraday depth maps according to Eq.~\ref{eq:Faraday_Depth_eq}. Taking into account possible depolarisation effects caused by the telescope beam finite size, we produced maps of the two key observables $\overline{\phi}$ and $\sigma_\phi$ of the Faraday depth spectrum and compare them with current radio observations. We have also computed radial profiles of the magnetic field components to probe the topology of the magnetic field in the final quiescent disc (see below). 

\subsection{Magnetic fields at high redshift}
\label{results:z=4}

\begin{figure*}
    \centering
    \includegraphics[width=1.0\textwidth]{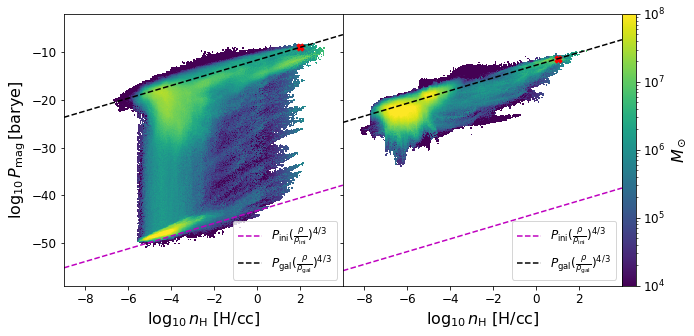}
    \caption{Mass weighted 2D histogram of magnetic pressure versus gas density at redshift $z=4$ (left) and $z=0.2$ (right) within 10 $R_\mathrm{vir}$ using quenching parameter $q = 10^3$. The two dashed lines show the magnetic pressure arising from compression alone ($P \propto \rho^{4/3}$), normalized to the initial field (bottom dashed line) and to the equipartition field (top dashed line). In each case, we show with a red cross the physical conditions at the center of the galactic disc that was used to normalize the upper scaling relation. }
    \label{fig:phase_digram_comparison}
\end{figure*}

\subsubsection{Magnetic phase space diagram}

In our simulation, the initial seed field strength was set to a value corresponding to the Biermann battery, namely $10^{-20}$~G at $z=10$.
This field then evolves due to expansion and gravitational collapse alone, according to the expected frozen-in behaviour of ideal MHD,  
following the scaling $P_{\rm mag} = P_\mathrm{ini} (\rho/\rho_\mathrm{ini})^{4/3}$. We plot in Fig.~\ref{fig:phase_digram_comparison} the mass-weighted 2D histogram of magnetic pressure versus gas density in a spherical region of radius 10~$R_\mathrm{vir}$, centered on the most massive halo at redshift $z = 4$ and $z = 0.2$. In the left panel, we see at $z=4$ a lot of ``pristine'' gas following the expected ideal MHD scaling. 
Once gas collapses and sets into a star forming galactic disc, the subgrid turbulent dynamo starts to amplify the field. 
The effect of the subgrid dynamo can be seen in Fig.~\ref{fig:phase_digram_comparison} as the different parallel tracks, each track corresponding to
a different progenitor galaxy at a different stage of its own galactic dynamo. 

The typical time scale for the subgrid dynamo is quite fast,
with an e-folding time scale equal to $\Delta x / \sigma_{\rm T}$. For a resolution of 100~pc and a turbulent velocity dispersion of $30$~km/s (typical of dwarf galaxies at that epoch), this gives a subgrid dynamo growth time scale of $3$~Myr, much faster than the Hubble time. 
We see in Fig.~\ref{fig:phase_digram_comparison} that these parallel tracks are restricted to a region of phase space to the right of the vertical line defined by 
$n_{\rm H}=10^{-2}$~H/cc, which is our density threshold for the subgrid dynamo. 

We finally show in Fig.~\ref{fig:phase_digram_comparison} as a black dashed
line the upper envelope of our magnetic phase space diagram. This relation corresponds to the same ideal MHD scaling as before,
but this time normalized to the equipartition field inside the dense ISM. To quantify further this relation, we have adopted
for the dense ISM typical values for the density $\rho_{\rm gal}$ and the magnetic field $P_\mathrm{gal}$ shown as the red cross.
We have checked that the magnetic pressure corresponds roughly to the equipartition value with the subgrid turbulence
where $P_{\rm gal} \simeq K_{\rm T}$. The ideal MHD scaling in the upper envelope can be computed using 
$P_{\rm mag} = P_\mathrm{gal} (\rho/\rho_\mathrm{gal})^{4/3}$. This scaling law can be used to estimate 
the magnetic field in the lower density gas of the galactic winds escaping the galaxies. Note that the higher magnetic field strength
observed in our simulation below $n_{\rm H}=10^{-2}$~H/cc is solely due to galactic winds, as we have no subgrid dynamo there.

The right panel of Fig.~\ref{fig:phase_digram_comparison} shows the same phase space diagram but this time at $z=0.2$.
At this late epoch, there is no pristine gas left within 10~$R_{\rm vir}$ of the main halo. 
The entire region has been contaminated by the equipartition field of the galaxy transported outside by galactic winds. 
Here again, we show the dense ISM typical values as the red cross, and from there we draw a dashed line following the ideal MHD scaling.
We see that this simple model describes quite well the upper envelope of the magnetic phase space diagram.
Note that the scaling relation at low redshift is almost two order of magnitude lower than the same scaling relation at high redshift, demonstrating that magnetic fields
in our simulation are stronger in the past than they are today.

\begin{figure*}
    \centering
    \includegraphics[width=1.0\textwidth]{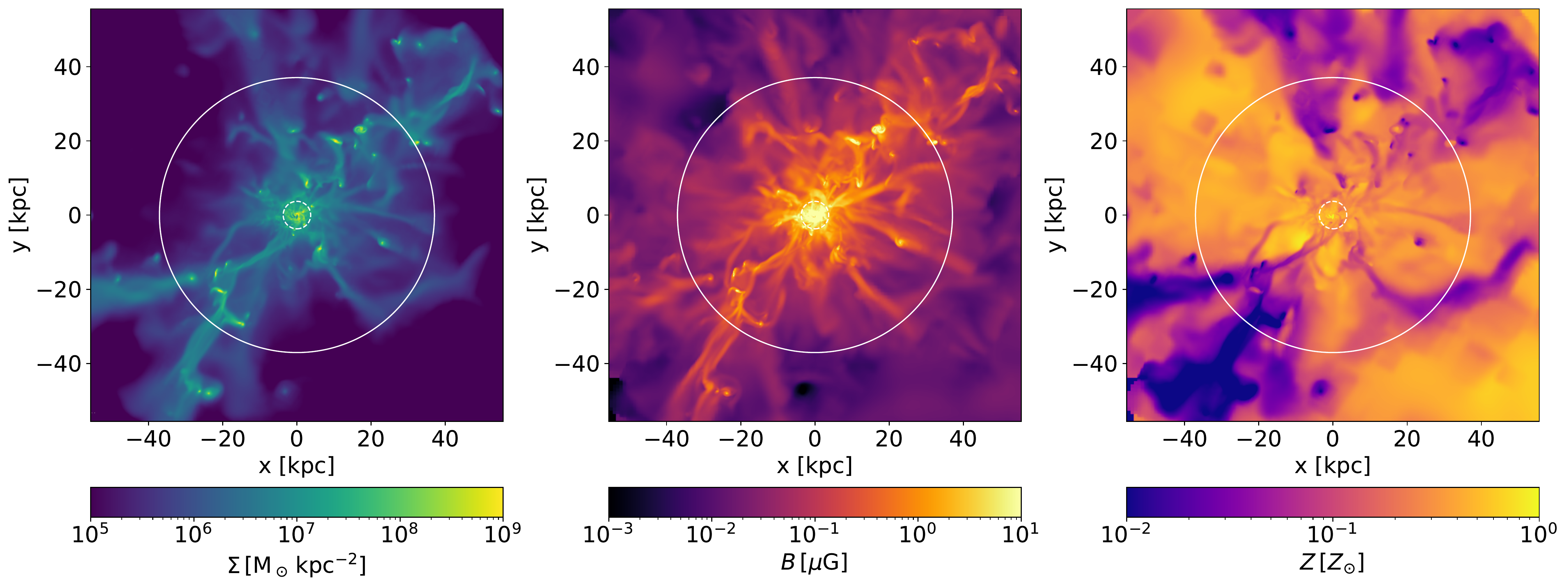}
    \caption{Maps of the gas surface density (left), the magnetic field strength (middle) and metallicity (right) in the circumgalactic region at redshift $z = 4$. The field strength and the metallicity are both density-weighted and averaged along the line of sight. The box is 100~kpc wide in the z-direction. The virial radius $R_\mathrm{vir}$ is indicated by the white circle and the galaxy radius 10\% $R_\mathrm{vir}$ is indicated by the dashed circle. }
    \label{fig:image_density_mag_pre_metal}
\end{figure*}

\subsubsection{Spatial distribution of the magnetic field}

We now analyze the spatial distribution of the magnetic field in and around our high-redshift galaxy. We show in Fig.~\ref{fig:image_density_mag_pre_metal} maps of the projected gas surface density, density-weighted magnetic field strength and density-weighted metallicity. The virial radius is shown as the solid circle, while the central galaxy is shown as the dashed circle. 
The field strength varies between 100~$\mu$G inside the galaxy to 1~$\mu$G outside in the halo. This is considerably more than for 
a typical low redshift galaxy, and consistent with the fact that at high-redshift, turbulence is high in the disc and outflows are strong in the halo.
The metallicity map confirms that most of the halo magnetic field comes from metal-enriched material in the outflows. We find good agreement between our results and the one published recently by \citet{Pakmor.2020}, although they did not use any subgrid model for the turbulent dynamo but started with a much higher initial seed field.

\begin{figure*}
    \centering
    \includegraphics[width=1.0\textwidth]{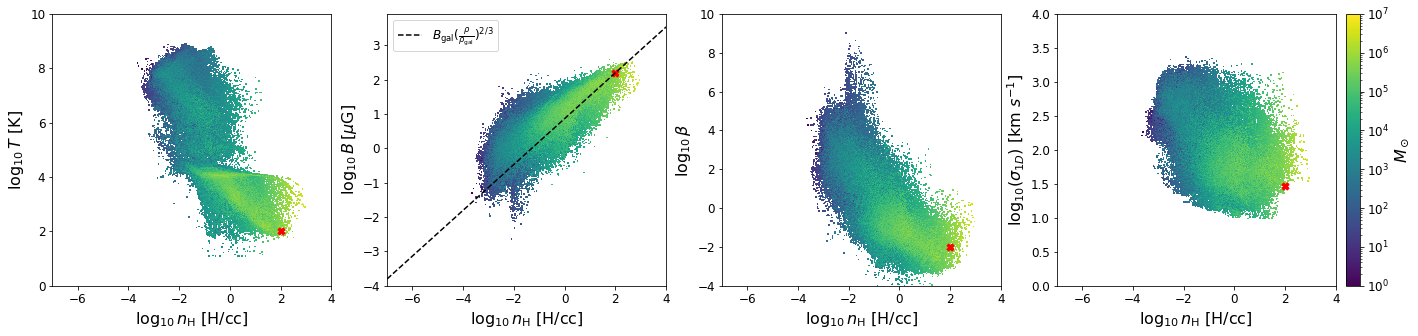}
    \caption{Mass-weighted 2D histogram of temperature, magnetic field strength, plasma $\beta$ and velocity dispersion as a function of gas density within 10\% $R_\mathrm{vir}$ at $z = 4$. The red crosses mark the typical values of these four variables given a density of 100 H/cc. The dashed line in the $B-\rho$ plot shows the scaling relation due to stretching and compression of the gas. }
    \label{fig:disk_variables_020}
\end{figure*}

In Fig.~\ref{fig:disk_variables_020}, we plot the temperature, magnetic field strength, plasma $\beta$, and turbulent velocity dispersion as a function of gas density within the central galaxy, defined as a sphere of radius $10\% R_\mathrm{vir}$. As before, we have added a red cross in each phase space diagram to pick up a typical value for each physical property. We also show in the magnetic phase space diagram a dashed line indicating the expected ideal MHD scaling relation $B \propto \rho^{2/3}$, normalized to the central values. The dense ISM shows a combination of low temperature (around 100~K) and high magnetic field strength (100~$\mu$G) leading to $\beta \simeq 0.01$. In the warm ISM, $\beta$ is closer to unity and becomes larger as one enters the low density and hot gas associated to the outflows. 

The typical physical conditions in the central region of our high-redshift galaxy, as marked by the red cross, have a gas number density of 100 H/cc, a gas temperature of 100 K and a plasma $\beta$ of $10^{-2}$. The typical magnetic field strength in this region, around 100 $\mu$G, gives a magnetic pressure around $10^{-9}$ barye. The typical one-dimensional velocity dispersion is about 30 km/s, which results in a turbulent energy density of roughly $2 \times 10^{-9}$ erg/cm$^3$. We conclude that in the central region, the magnetic field is slightly below equipartition (close to 0.5) and way above the quenching parameter controlling the saturation of our subgrid dynamo. 

\subsubsection{Comparison to Faraday depth observations}

\begin{figure*}
    \centering
    \includegraphics[width=0.8\textwidth]{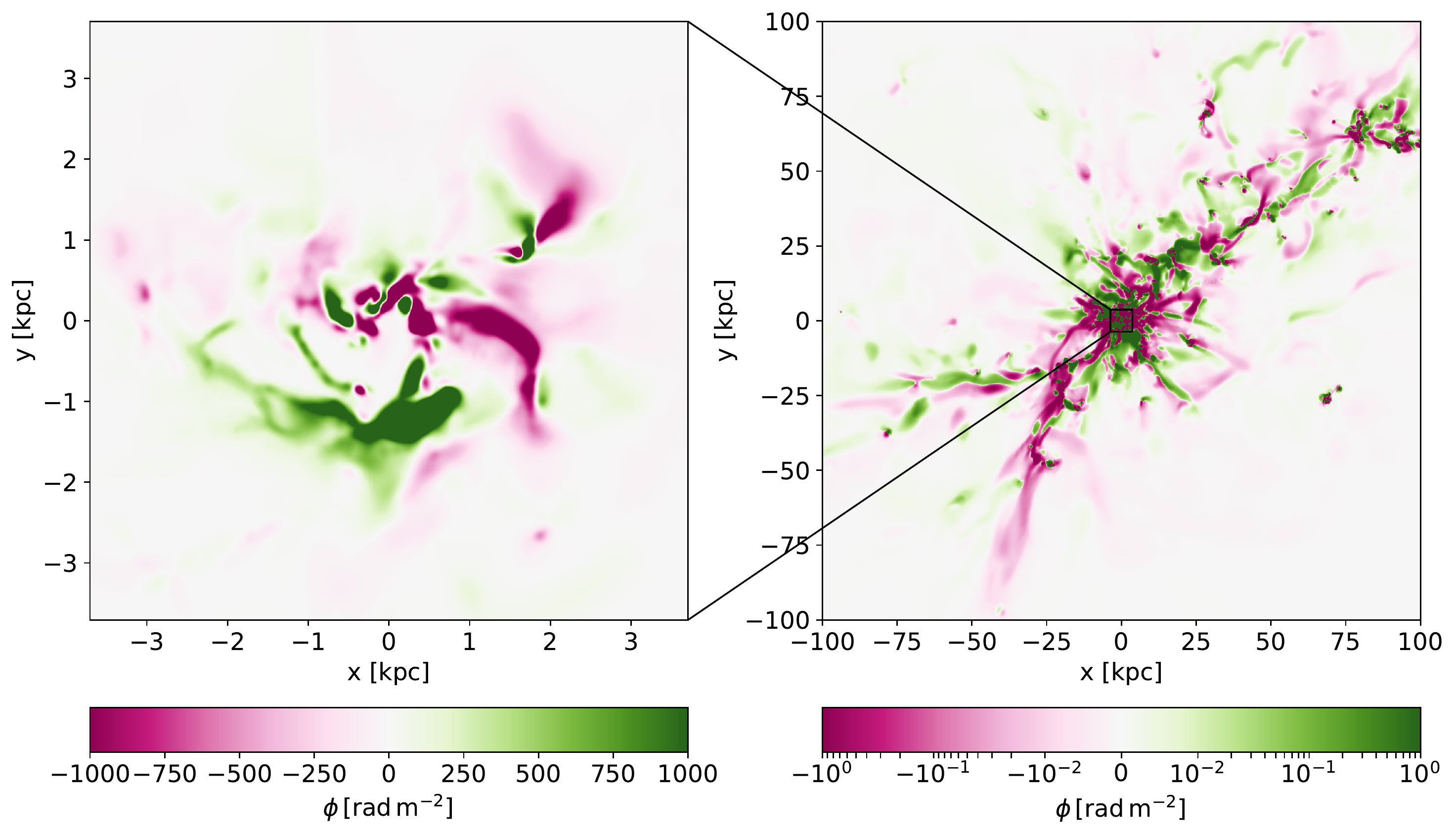}
    \caption{Faraday depth map at z = 4 as seen by an observer. The left panel shows the zoom-in of the Faraday depth map of the halo on the right. }
    \label{fig:Faraday_depth_a_020}
\end{figure*}

Assuming that our simulated galaxy is located between an observer and a distant quasar, the polarised electromagnetic emission of the quasar will be affected by the magnetic field along the line of sight. 
Using Eq.~\ref{eq:Faraday_Depth_eq}, we produce the Faraday depth maps shown in Fig.~\ref{fig:Faraday_depth_a_020}. Note that the field of view in the right panel is more extended than in Fig.~\ref{fig:image_density_mag_pre_metal} to show the cold streams connecting the halo to the cosmic web. The Faraday depth in these filaments is quite low, slightly below $0.1$~rad~m$^{-2}$, while it rises to values larger than $1$~rad~m$^{-2}$ in the inner halo. The left panel shows a zoom-in on the central galaxy, where the Faraday depth is quite large, going beyond 1000~rad~m$^{-2}$ in some places. In this image, the magnetic field appears as very turbulent, without a clear large scale signal for the Faraday depth. This means that at this redshift, the large scale magnetic field is negligible compared to the small scale random turbulent field. 

\begin{figure*}
    \centering
    \includegraphics[width=0.8\textwidth]{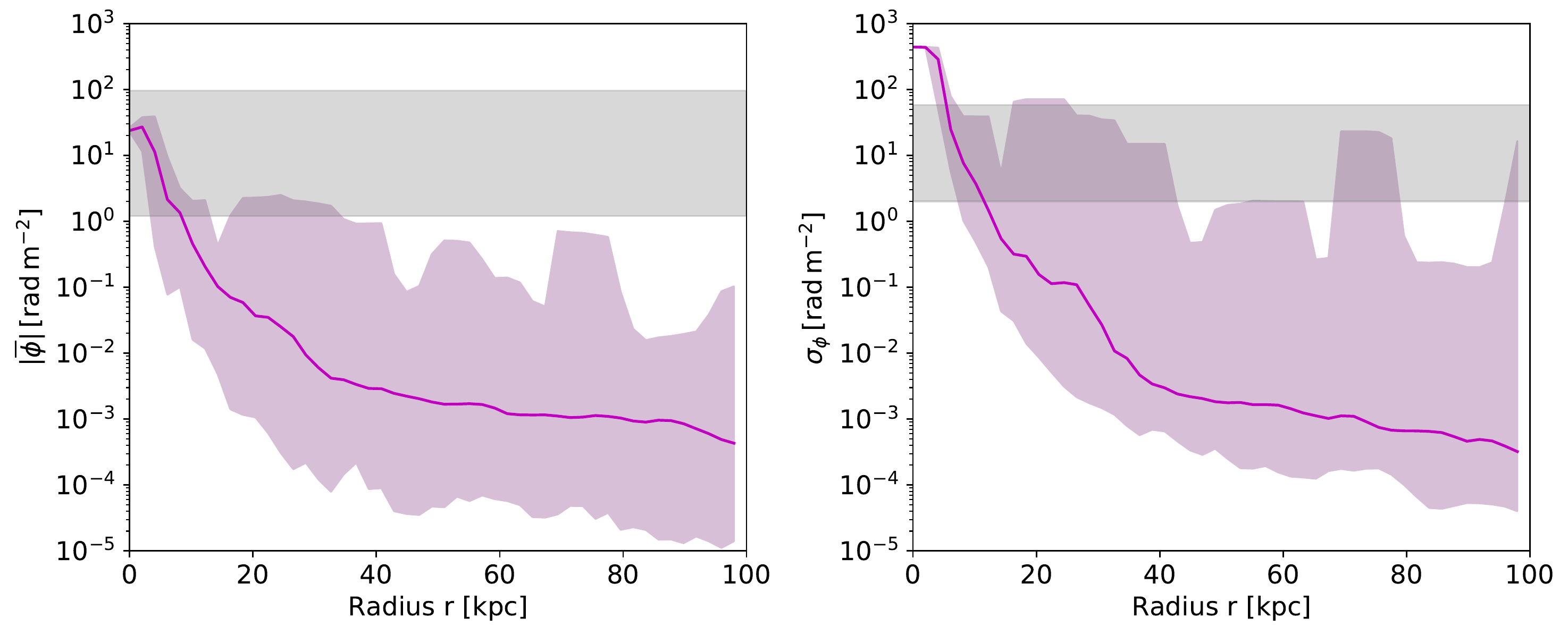}
    \caption{Radial profiles of $\overline{\phi}$ and $\sigma_{\phi}$ averaged within annuli of radius $r$ and a thickness of 2 kpc. The purple curve indicates the median values within the annuli. The purple shaded area covers 97\% of all the data and the grey shaded area show the values observed by \citet{Kim.2016dx2}. }
    \label{fig:Faraday_profile}
\end{figure*}

As discussed in Sec.~\ref{sec:Faraday}, we have modelled telescope beam finite size effects, which can lead to a significant Faraday depolarisation in presence of such turbulent fields. The Very Large Array has a maximal angular resolution of 2~arcsec in the L~band \citep[][]{1980ApJS...44..151T}. This corresponds to a spatial resolution of around 10 kpc at $z=4$. Hence, each pixel of the Faraday depth map in the right panel of Fig.~\ref{fig:Faraday_depth_a_020} has been smoothed over a circular region of 10 kpc in diameter. Within this circular region, we compute the mean Faraday depth $\overline{\phi}$ and the standard deviation $\sigma_\mathrm{\phi}$ for each image pixel. The radial profiles of $\overline{\phi}$ and $\sigma_{\phi}$ are shown in Fig.~\ref{fig:Faraday_profile}. We see that the variance of the Faraday depth is almost everywhere one order of magnitude larger than the mean, a clear signature of a small scale turbulent field for which depolarisation is quite strong. 

We would like to compare to the observations of \citet{Kim.2016dx2}, for which Faraday spectra of 49 unresolved quasars have been collected and analysed using Faraday synthesis. They found that strong depolarisation is correlated with strong Mg II absorption lines, which is considered as a clear signature of outflows in the inner halo. Our simulated galaxy supports this picture of highly turbulent star forming galaxies driving strong outflows and powering a saturated turbulent dynamo. The observed depolarisation is the smoking gun for a turbulent magnetic field. 

More quantitatively, we show in Fig.~\ref{fig:Faraday_profile} as a grey region the range of the observed values for the mean and the variance of the ``in beam'' Faraday depths. Although we see a reasonable agreement between our  predictions and the observations, our simulated galaxy produces Faraday depths that are systematically lower than the one measured in this sample of 49 quasars. The main caveat in our current analysis is that the observed halos cover a large range of redshifts and virial masses that do not match perfectly our single simulated halo at $z=4$.  This would require a larger simulation suite with multiple halo masses
that goes beyond the scope of this paper.

\subsection{Magnetic Fields at low redshift}
\label{results:z=0.2}

After the last (major merger driven) starburst around $z \simeq 1.5$, the turbulent galactic environment settles into a quiescent, disc-dominated phase.
As a consequence of the lower gas surface density, stellar feedback is not efficient anymore at launching outflows and a razor thin disc configuration emerges. 
The sequence of events leading to the formation of this extended disc is described in great detail in \cite{Kretschmer.2020}. The final state of our simulated galaxy turns out to be similar to many large spiral galaxies found in the nearby Universe.

\begin{figure*}
    \centering
    \includegraphics[width=0.8\textwidth]{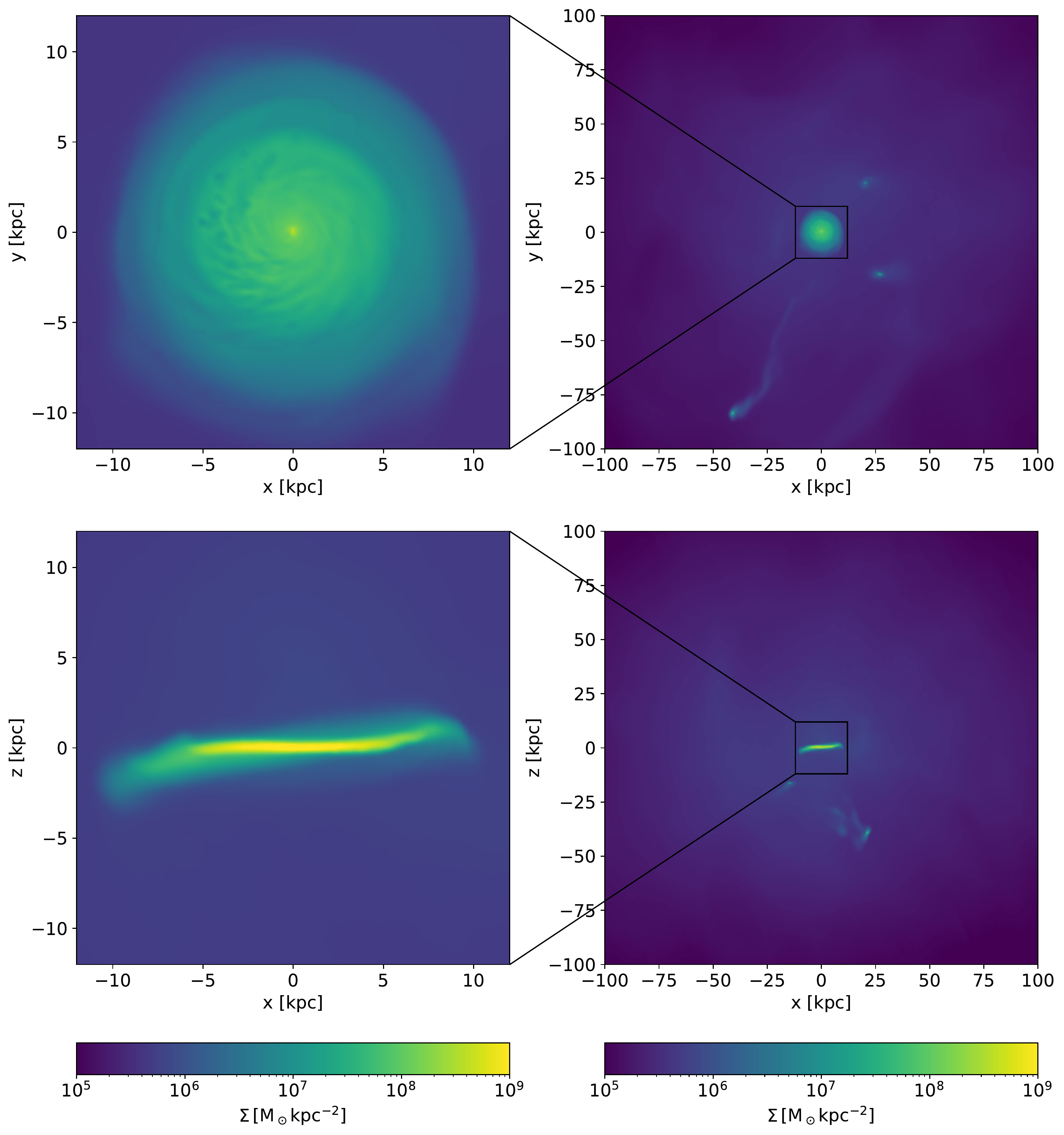}
    \caption{Gas density integrated along the line of sight at redshift $z=0.2$. The top row shows the face-on views of the disc and the halo while the bottom row shows edge-on views.}
    \label{fig:gas_density_a_079}
\end{figure*}

We show in Fig.~\ref{fig:gas_density_a_079} the gas surface density with the disc face-on and side-on, both at the halo scale for the right panels and at the central galaxy scale on the left panels. The side-on view shows that the gas disc is particularly thin and barely resolved by our smallest 55~pc AMR cells. The face-on view reveals a weak spiral within a floculent disc. The disc looks very isolated at this late epoch, with only a few low mass satellites orbiting the halo. The gas in the halo is very homogeneous, with only a few ram pressure stripped tails following the satellites. 

\begin{figure*}
    \centering
    \includegraphics[width=0.8\textwidth]{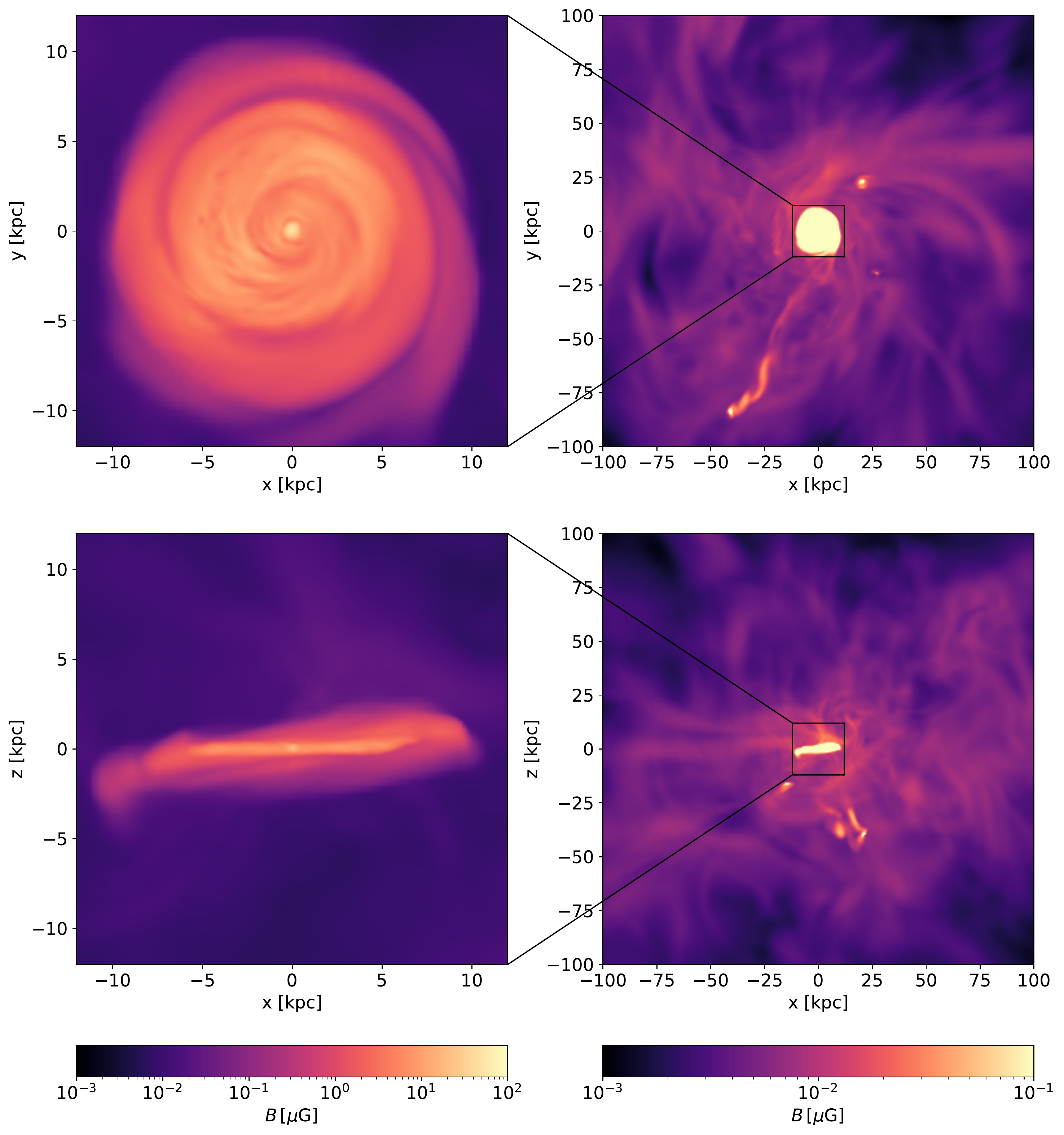}
    \caption{Averaged magnetic field strength weighted by gas density along the line of sight at redshift $z=0.2$. The top row shows the face-on views of the disc and the halo while the bottom row shows edge-on views.}
    \label{fig:mag_norm_a_079}
\end{figure*}

The companion plots in Fig.~\ref{fig:mag_norm_a_079} show the magnetic field strength along the line of sight, weighted by density. The field is about 10 $\mu$G in the disc and 100 $\mu$G in the nuclear region, in good agreement with radio observations of the magnetic field in the Milky Way that are consistent with an average field strength of 10 $\mu$G  and a much higher field strength towards the Galactic centre \citep{2013pss5.book..641B}. In the circumgalactic medium, we see a weaker field of the order of tens of nG.  In their AURIGA simulations,  \cite{Pakmor.2020} have also obtained magnetic fields ranging from a few $\mu$G to tens of $\mu$G inside the galactic discs and around $10$~nG in the halo. Using an analytical approach, \cite{Beck.2012} designed a simple model for the magnetic field in the Milky Way halo, with a field strength of about 1~$\mu$G in the centre down to 1~nG in the intergalactic medium, in good agreement with our numerical results.

\begin{figure*}
    \centering
    \includegraphics[width=1.0\textwidth]{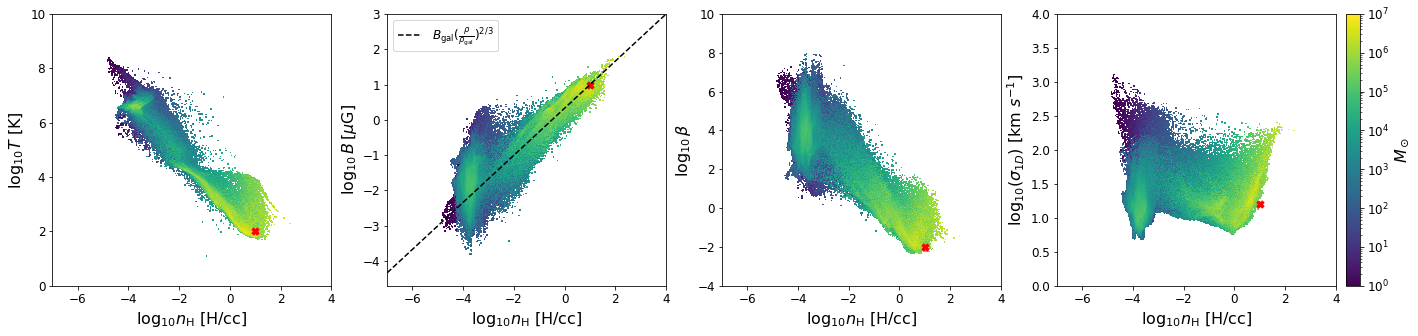}
    \caption{Mass-weighted phase space diagrams of temperature, magnetic field strength, plasma $\beta$ and velocity dispersion as a function of gas density 
    within 10\% $R_\mathrm{vir}$ at $z=0.2$. The red crosses mark the typical values of these four variables in the central region of the disc. 
    The dashed line shows the ideal MHD $B \propto \rho^{2/3}$ scaling relation, normalised to the galactic central region, where the field is in equipartition 
    with the subgrid turbulence. }
    \label{fig:disk_variable_079}
\end{figure*}

In Fig.~\ref{fig:disk_variable_079}, we show various phase space diagrams such as the temperature, the magnetic field strength, the plasma $\beta$ and the subgrid turbulent velocity dispersion versus the gas density in the galaxy (defined as a sphere of radius 10\% $R_\mathrm{vir}$). In the central region of the disc, the gas is cold and dense with a temperature of about 100~K and a density of about 10~H/cc, as shown by the red cross in each panel. The velocity dispersion of the unresolved subgrid turbulence is also quite small, around 10~km/s, typical of Milky Way like galaxies. Note that the smooth appearance of our gas disc suggests that our simulation does not have enough resolution to capture any resolved turbulence in the disc, contrary to its high redshift counterpart. The magnetic field in the central region peaks at about 20~$\mu$G (see the red cross in the figure), resulting in a low plasma $\beta \simeq 10^{-2}$ and a peak magnetic pressure of about  $2 \times 10^{-11}$~barye, in rough equipartition with the subgrid turbulent pressure. This demonstrates again that the magnetic field is getting lower at lower redshift because the turbulence in the disc is decreasing, as the galaxy settles in a more quiescent mode. Note also that the final field strength is much larger than the value required to quench our subgrid dynamo, so that at late time our subgrid turbulent dynamo is totally ineffective, as it should.

\begin{figure*}
    \centering
    \includegraphics[width=0.8\textwidth]{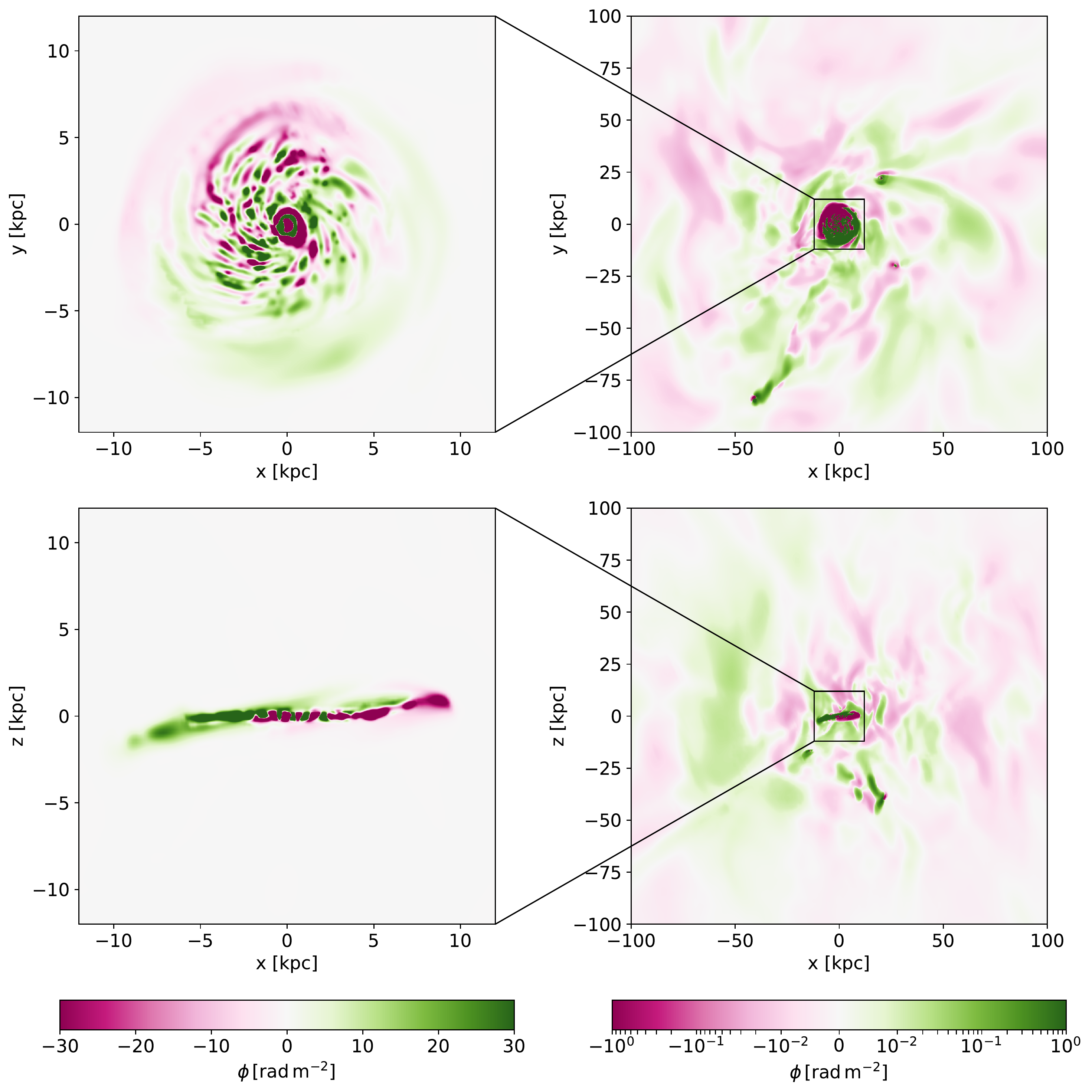}
    \caption{Faraday Depth map of the zoomed region at z = 0.2 as seen by an observer. The two rows show face-on and edge-on projected Faraday Depth maps of the simulated galaxy. The left column shows zoom-ins of face-on and edge-on Faraday Depth maps of the halo. }
    \label{fig:Faraday_depth_a_079}
\end{figure*}

We also produce mock observations of Faraday depth maps at $z=0.2$, shown in Fig.~\ref{fig:Faraday_depth_a_079}. These can be directly compared to the theoretical predictions of \citet{2018MNRAS.481.4410P}, showing good agreement between the two sets of simulations with a typical Faraday depth of 20~rad~m$^{-2}$ for a face-on view of the disc. Faraday depth are much weaker in the halo with values of the order of $10^{-2}$~rad~m$^{-2}$. 

Since the Faraday depth is sensitive to the magnetic field parallel to the line of sight, the face-on view reveals a very turbulent field typical of the vertical component $B_z$, while the side-on view shows a much more prominent large scale field typical of the dominant toroidal component $B_\phi$. 
A detailed study of M51 shows similar magnetic field strengths and sign reversals on small scales \citep{2011MNRAS.412.2396F}. In the circumgalactic region, the magnetic field fluctuations appear on  much larger scales. However, since the AMR resolution is quite degraded outside the galactic disc, we probably underestimate the amplitude of the small-scale fluctuations. \citet{2018MNRAS.481.4410P} also found Faraday depth of the order of 0.1~rad/m$^2$ in the circumgalactic medium, in good agreement with what we find here. 

\begin{figure*}
    \centering
    \includegraphics[width=1.0\textwidth]{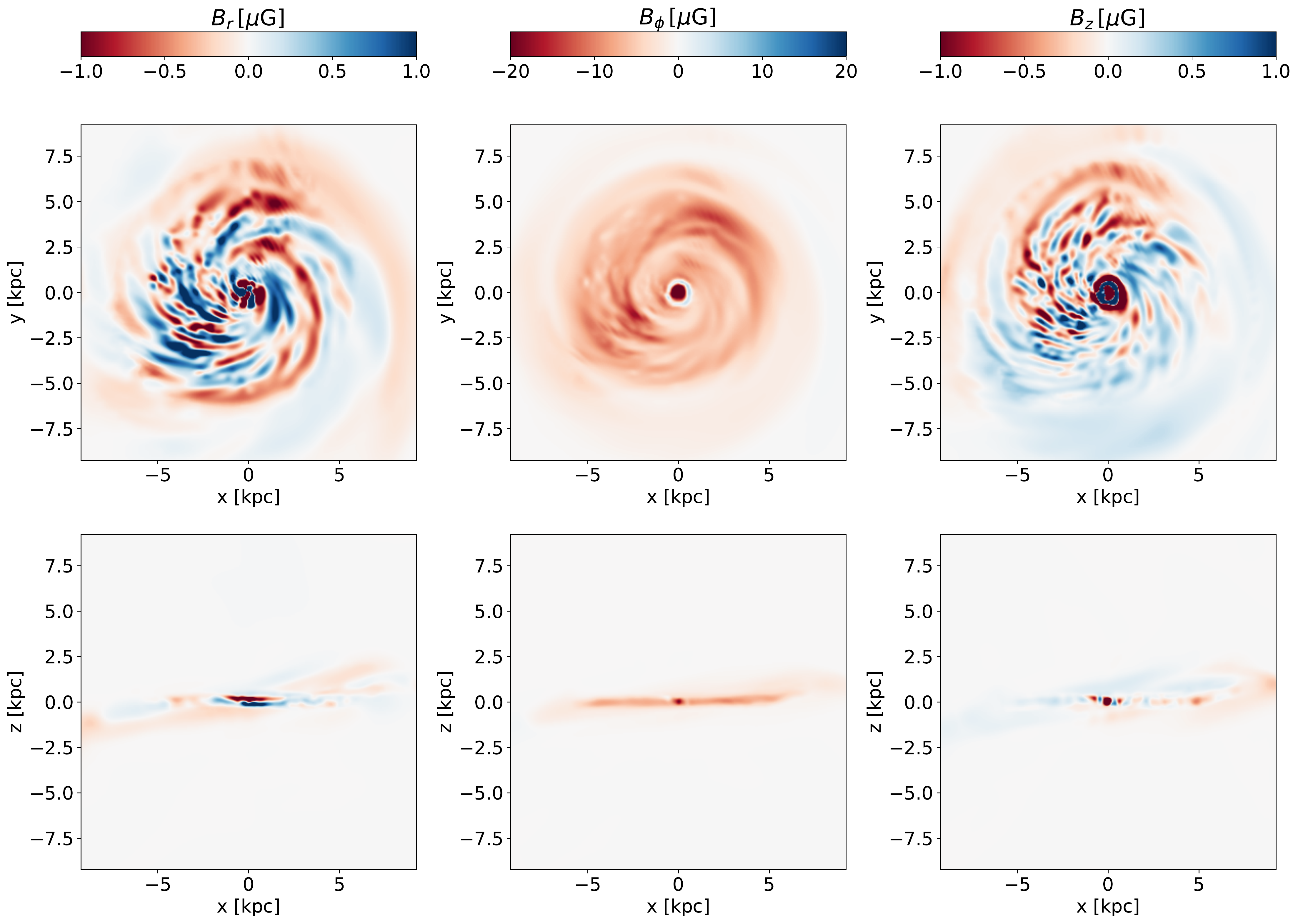}
    \caption{Magnetic field topology at $z=2$. The face-on (top row) and edge-on(bottom row) density-weighted projections of the magnetic field components in cylindrical coordinates. The columns show the radial field strength (left column), toroidal field strength (middle column), and the vertical field strength (right column), respectively. }
    \label{pic:field_topology_img}
\end{figure*}

The topology of the magnetic field in the disc can be understood by plotting each magnetic field component in cylindrical coordinates. Fig.~\ref{pic:field_topology_img} shows that the magnetic field is dominated by the toroidal component with a typical strength of around 10 $\mu$G. The orientation of the toroidal field is strikingly the same across the disc. The field strength of the vertical and the radial components is much weaker than the one of the toroidal field. The vertical and radial components also exhibit small-scale structures, with the field orientation flipping multiple times at a given radius. We don't see field reversals in the toroidal field at large radii, in disagreement with the findings of \citet{2018MNRAS.481.4410P}. We only see one toroidal field reversal around the nuclear region. The vertical and radial components flip multiple times across the mid-plane, while the toroidal field always aligns with the same direction. We believe these results might be due to the absence of resolved turbulence in the disc, leading to an abnormally quiescent flow. This almost perfectly rotating disc leads to a large-scale toroidal field significantly larger than both the vertical and the radial component. A better spatial resolution would have triggered more random motions, especially in the radial direction, leading to more field reversals and a larger pitch angles for the field.

\begin{figure*}
    \centering
    \includegraphics[width=1.0\textwidth]{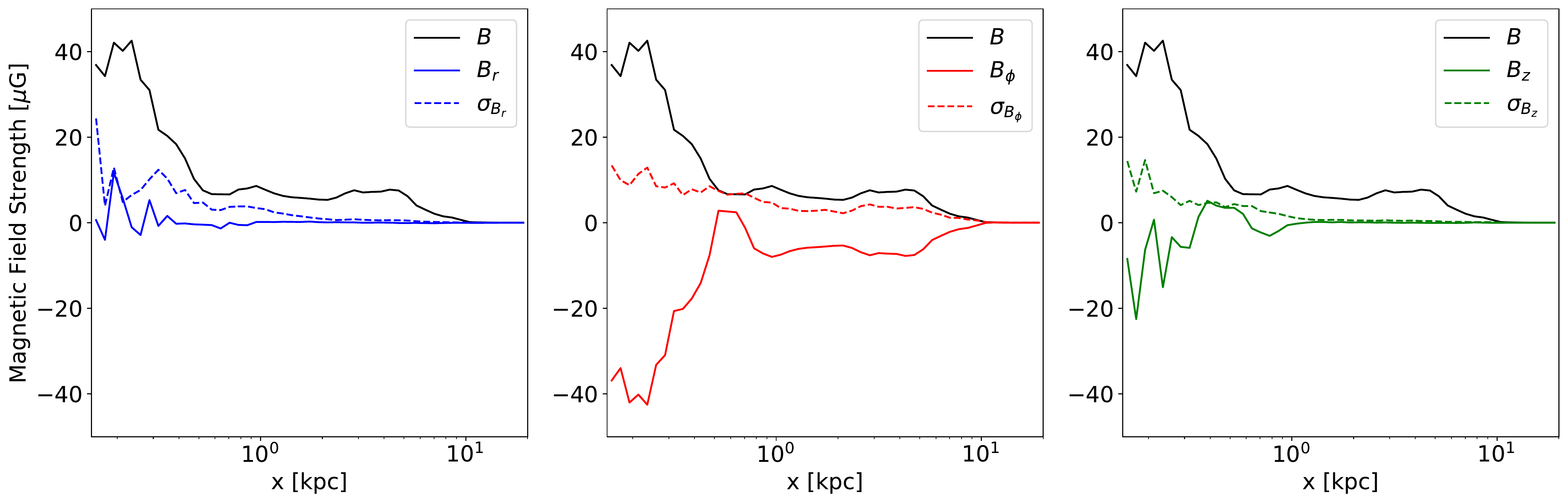}
    \caption{Average magnetic field strength and standard deviation of the radial, toroidal and vertical components in cylindrical shells as a function of radius. The total field strength is plotted in black. }
    \label{pic:mag_field_cylin_profile}
\end{figure*}

Fig.~\ref{pic:mag_field_cylin_profile} further supports this conclusion, showing radial profiles for each magnetic field component. Both the angle-averaged field strength and the standard deviation are calculated in cylindrical shells of thickness 1~kpc. The total magnetic field is clearly dominated by the toroidal component, for which the large-scale contribution dominates over the small-scale fluctuations captured by the variance. One can also see that the toroidal field has the same orientation as seen in Fig.~\ref{pic:field_topology_img}, except one reversal at $r=0.5$ kpc, which is located outside the nuclear disc. For both the radial and vertical components, the standard deviation is in general larger than the angle-averaged value, which indicates that small-scale fluctuations dominate over the large-scale average value. 

\section{Discussions}
\label{sec:discussion}

In this paper, we have performed a cosmological zoom-in simulation of a Milky Way analogue, modifying the induction equation using a mean-field approach with a simple subgrid turbulent dynamo model. In an earlier series of papers \citep{Rieder.2016,Rieder.2017y2f,Rieder.2017}, we have studied small-scale turbulent dynamo amplification of a very small seed field driven by supernova feedback. These simulations were based on the unmodified induction equation (Eq.~\ref{eq:MHD_4}). We showed that in order to see a fast turbulent dynamo in a dwarf galaxy, the spatial resolution has to be better than $\Delta x \simeq 10 \mathrm{pc}$ \citep{Rieder.2016,Rieder.2017}. This resolution requirement is computationally prohibitive for simulations of large galaxies like the Milky Way, not to mention entire periodic boxes. In this paper, we have demonstrated that our subgrid model allows to capture a fast amplification of the field driven by unresolved turbulence on a few Myr time scale. This subgrid dynamo remains active until the magnetic energy density reaches a small fraction (typically here $q \simeq 10^{-3}$) of the subgrid turbulent kinetic energy, while the resolved turbulent and rotational motions further amplified the field up to equipartition.

In our model, the subgrid dynamo is restricted to dense gas ($n_{\rm H}>10^{-2}$~H/cc) defining the dense and supersonic  turbulent ISM. This means that magnetic fields in the IGM can only be of primordial origin (in our picture the Biermann battery seed fields) or polluted by galactic winds. This is inline with the model proposed by \citet{2006MNRAS.370..319B}. In their semi-analytical model, the galactic winds are never entirely volume-filling. Pockets of pristine gas in cosmic voids unaffected by feedback processes at $z=0$ remain. In our zoom-in simulation, by $z=0$, the entire region within 10 $R_\mathrm{vir}$ of our Milky way-like galaxy has been polluted by galactic winds with magnetic fields everywhere of the order of 1~nG. We have not found any pocket of pristine gas with a magnetic field strength around $10^{-20}$~G within this region. A definitive answer would probably come from a larger periodic box simulation, although resolving the formation of the entire range of halo masses, down to star-forming mini-halos would prove particularly challenging.

In an attempt to compare to observations, we have computed the Faraday depth of our halo at $z=4$, creating mocks of the observations performed in~\citet{Kim.2016dx2}. These are interferometric observations that probe the polarization properties of background quasars. As discussed in Sec.~\ref{results:z=4}, in our simulation, the CGM is filled with highly turbulent and strongly magnetised gas powered by galactic outflows. Although the level of depolarisation in our simulation agrees qualitatively well with the observations, we predict a weaker overall Faraday depth. The main caveat here is that we only consider a single galaxy, while observations span a large range of redshifts and halo masses. A larger sample of simulated halo masses is therefore required and will be the topic of a follow-up paper. We can also compare to other theoretical predictions with similar halo masses, such as the work of \citet{2018MNRAS.481.4410P}. The agreement is quite good with this AURIGA sample of Milky Way analogues, both inside the star forming galaxy and in the CGM around the galaxy. This supports that the theoretical picture that emerges is robust: first, a saturated turbulent dynamo augmented by large-scale rotation amplifies the magnetic field close to equipartition and second, galactic outflows transport this equipartition field to the outskirts of the galaxies and in the IGM. These two very different regions are connected to each other by the ideal MHD scaling relation $B \propto \rho^{2/3}$.

At low redshift, the situation is radically different. The galaxy ends up in a much more isolated environment. It features a large and thin disc with weaker star formation and no visible outflows. 
The subgrid (unresolved) turbulence is much weaker, as well as the resolved turbulence. The disc is so thin that it is not clear if we are resolving any turbulent vertical motion at all. As a result of this dramatic evolution to this low redshift quiescent state, the magnetic field in and around the galaxy is much weaker than in its high redshift progenitor, with a field strength of about 10~$\mu$G in the disc and 10~nG in the halo. This is in good agreement with the simulation of  \citet{Pakmor.2017} and \citet{Pakmor.2020}, with some interesting differences in the details that probably trace the different numerical methodologies. In particular, our magnetic field is a factor of 2 to 4 weaker than in \citet{Pakmor.2020}, possibly a consequence of our less turbulent halo gas.

We found that the magnetic field in the final disc is mostly toroidal, a consequence of our resolved rotational motions in the late time evolution of the spiral galaxy. The lack of resolved turbulent motions in our final razor-thin state probably leads to an underestimation of the amplitude of the radial and vertical components of the field. Only better resolved simulations will confirm this. 

The very simple (probably simplistic) form we have adopted for our $\alpha$ tensor (see Eq.~\ref{eq:alpha_expression}) could be improved in many ways. One could estimate the local, mean-field helicity and uses it to provide a more realistic form to the pseudo-scalar $\alpha$. A more ambitious goal would be to derive a new mean-field equation for $\alpha$ with source and sink terms governing its evolution \citep[see][for a review of possible methods]{Brandenburg.2005}. It is unclear whether these possible refinements in the theory would significantly change our results. We believe that the most important aspect is to resolve properly the turbulent motions on large scales (meaning here the grid scale and above). We cannot claim we achieved this at low redshift, as discussed already multiple times, but we certainly succeeded to fulfil this requirement at high-redshift. 

Another caveat of our model is how we set up the initial magnetic field. We used a uniform field with a strength of $10^{-20}$~G, mimicking the outcome of the Biermann battery during the epoch of reionization. The topology of the Biermann generated fields is far more complex \citep[see][for a recent study of the process]{Attia.2021}. In the context of the subgrid turbulent dynamo, the topology and coherence length of the initial field is particularly important. Indeed, since in our simple model $\alpha$ is a true positive scalar, the subgrid turbulent dynamo amplifies the initial field parallel to itself. Any initial random orientations will be preserved during the amplification process. In principle, resolved turbulent motions will randomize the field close to the grid scale and change its direction accordingly. Nevertheless, exploring the effect of different initial field topology and coherence lengths would clearly justify future studies. 

\section{Conclusion}

We have designed a new subgrid turbulent dynamo numerical scheme for galaxy formation simulations. It allows us to exponentially amplify an initially very small field to equipartition strength at both low and high redshift, without having to either rely on a very high and demanding spatial resolution or use unrealistically large initial field strength values. 

This subgrid dynamo exploits a subgrid scale (SGS) mean-field turbulence model that describes creation and dissipation of turbulence below the grid scale of the supersonic unresolved ISM. 

The $\alpha$ term that enters the modified induction equation is assumed to be proportional to the velocity dispersion of this unresolved turbulence, and to a quenching term that model the saturation of the subgrid dynamo. This quenching depends on one free parameter, namely the quenching parameter $q$. This parameter has been calibrated to a fiducial value of $q=10^{-3}$ that must be adjusted when changing the resolution. 

These simulations support the simple picture of a strong magnetic field in rough equipartition with the turbulent kinetic energy inside galaxies and strong galactic outflows pushing the field outside of the galaxy all the way into the IGM. 

This gives rise to field strength around 10~$\mu$G in low redshift quiescent discs and 100~$\mu$G in high redshift galaxies. The host halos contain a weaker field, about 10~nG at low redshift up to 1~$\mu$G at high redshift. These values are consistent with observations both at low and high redshift. For the latter, strong depolarisation of the Faraday signal is expected due to turbulence.

Both the SGS subgrid model for turbulence and the subgrid dynamo are available in the public version of the RAMSES code. 

We plan to perform other simulations in the near future to test and improve further our model and explore its predictions at smaller scales with idealised simulations of isolated galaxies and at larger scales with simulations of galaxy clusters.

\section*{Acknowledgements}
The authors thank the referee for the constructive comments. We thank Simon Lilly for helpful discussions on the complex topic of Faraday synthesis. We also thank Ulrich Steinwandel for helpful discussions and for comments on an early version of the manuscript. This work was supported by the Swiss National Supercomputing Center (CSCS) project s1006 - ``Predictive models of galaxy formation’' and the Swiss National Science Foundation (SNSF) project 172535 - ``Multi-scale multi-physics models of galaxy formation’’. The simulations in this work were performed on Piz Daint at the Swiss Supercomputing Center (CSCS, Lugano, Switzerland) and the analysis was performed with equipment maintained by the Service and Support for Science IT, University of Zurich.
We also made use of the Pynbody package \citep{pynbody}. 

%%%%%%%%%%%%%%%%%%%%%%%%%%%%%%%%%%%%%%%%%%%%%%%%%%
\section*{Data Availability}

The data underlying this article will be shared on reasonable request to the corresponding author.

%%%%%%%%%%%%%%%%%%%% REFERENCES %%%%%%%%%%%%%%%%%%

% The best way to enter references is to use BibTeX:

\bibliographystyle{mnras}
\bibliography{example,romain} % if your bibtex file is called example.bib
%
%
% Alternatively you could enter them by hand, like this:
% This method is tedious and prone to error if you have lots of references
%\begin{thebibliography}{99}
%\bibitem[\protect\citeauthoryear{Author}{2012}]{Author2012}
%Author A.~N., 2013, Journal of Improbable Astronomy, 1, 1
%\bibitem[\protect\citeauthoryear{Others}{2013}]{Others2013}
%Others S., 2012, Journal of Interesting Stuff, 17, 198
%\end{thebibliography}

%%%%%%%%%%%%%%%%%%%%%%%%%%%%%%%%%%%%%%%%%%%%%%%%%%

%%%%%%%%%%%%%%%%% APPENDICES %%%%%%%%%%%%%%%%%%%%%

%\appendix

%\section{Some extra material}

%If you want to present additional material which would interrupt the flow of the main paper,
%it can be placed in an Appendix which appears after the list of references.

%%%%%%%%%%%%%%%%%%%%%%%%%%%%%%%%%%%%%%%%%%%%%%%%%%

% Don't change these lines
\bsp	% typesetting comment
\label{lastpage}
\end{document}